# Effect of photons on atoms in crystallization of amorphous silicon films

NuoFu Chen[1,2*], Quanli Tao[1], Yiming Bai[1], Shaolin Ruan[2], Jikun Chen[3*]

1School of Renewable Energy, North China Electric Power University, Beijing 102206, China.

2Changzhou Yingnuo Energy Technology Co. Ltd., Changzhou, Jiangsu 213000, China.

3University of Science & Technology Beijing, Beijing 100083, China

*Corresponding author. Email: nfchen@ncepu.edu.cn

† Corresponding author. Email: jikunchen@ustb.edu.cn

**Abstract**：The effect of photons on atoms is revealed in the crystallization of amorphous silicon films. The crystallization of amorphous silicon films by conventional thermal annealing behaves (111) orientation superior. While a rapid thermal annealing (RTA) reveals that (220) orientation becomes stronger, and the other orientations become weaker. If RTA duration is long enough, except (220) orientation, all the other orientations disappear in x-ray diffraction spectra. The effect of photons on atoms take charge of the (220) preferred orientation. The mechanism of changes in orientations during the crystallization is analyzed in this paper.

**Introduction**

Due to their dependability and high power conversion efficiency, crystalline silicon solar cells are the most popular photovoltaic devices, controlling more than 90% of the market shares.[1-7] Amorphous silicon film solar cells were expected to be replacements for crystalline silicon solar cells because a lot of silicon materials and energy can be saved in the production of amorphous silicon film solar cells, and the products do not contain any hazardous materials. However, there is a fatal flaw, light-induced degradation, in amorphous silicon film solar cells,[8,9] which has not been well understood and, of cause, has been kept unsolved. The purpose of crystallizing amorphous silicon films is to reduce the light-induced degradation in amorphous silicon film solar cells.[10-12] There are several methods could be used to crystallize amorphous silicon films, such as conventional thermal annealing (CTA) under high temperature,[13,14] rapid thermal annealing (RTA),[15-17] metal induced crystallization,[18-20] and laser annealing.[21-24] Because the specific free surface energy of (111) plane is the smallest among all the other crystalline planes,[25,26] the crystallization of amorphous silicon films by conventional thermal annealing behaves (111) orientation superior. While rapid thermal annealing (RTA) reveals that (220) orientation becomes stronger, and the other orientations become weaker with time extending. If RTA duration is long enough, except (220) orientation, all the other orientations disappear in x-ray diffraction spectra. It is still an unsolved mystery that how are orientations changing during the crystallization of amorphous silicon films by RTA method. We show the interesting results of crystallization in amorphous silicon films by RTA, explore the process of the crystallization, and finally reveal the mechanism of changes in orientations during the crystallization in this paper.

**Experimental Section**

**Magnetron sputtering.** The amorphous silicon films are deposited about 5 microns thick using a self-designed magnetron sputterer on graphite wafers of 50 mm long, 50 mm wide, and 0.5 mm thick, from a p-type poly-crystalline silicon target at 600°C of substrate temperature. The deposition speed of amorphous silicon films is related to sputtering powers. The deposition speed in our experiments can be expressed by Equation (1) and (2) for the sputtering powers of 200W and 300W at 600°C of substrate temperature, respectively.

$$d = 0.447t + 0.00638t^2 \tag{1}$$

$$d = 1.314t + 0.0172t^2 \tag{2}$$

Where, $t$ and $d$ are the sputtering time and thickness of amorphous silicon films respectively.

**Rapid thermal annealing.** It is closely related with the temperature of the samples which reflects the colour temperature of the tungsten-halogen lamps to crystallize amorphous silicon films. The sketch of rapid thermal annealing furnace used in our experiments is shown in Figure 1.

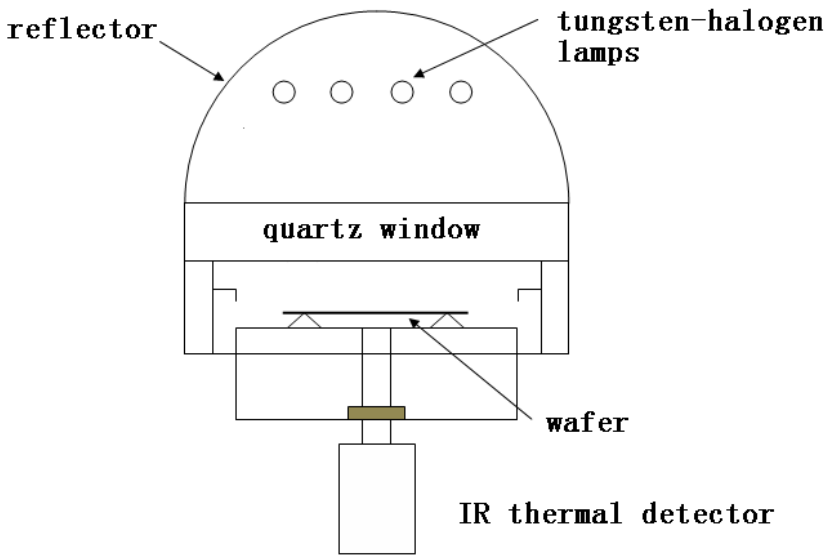

**Figure 1.** Sketch of rapid thermal annealing furnace

## Results and Discussion

**Crystallization of amorphous silicon films**

X-ray diffraction (XRD) measurements show that the sputtered silicon films deposited by magnetron sputtering technique are amorphous because only diffraction peaks from graphite substrates can be seen, as shown in Figure 2(A). Then the amorphous silicon films are crystallized by CTA with a tube furnace and RTA with tungsten-halogen lamps, respectively. The results of XRD measurements indicate that (1) the amorphous silicon films can be crystallized with (111) superior orientation by CTA taking time more than several hours, under 1100℃, and Figure 2(B) shows a CTA result of 50 hours; (2) the amorphous silicon films can also be crystallized by RTA taking only dozens of seconds, under 1100℃, and (220) orientation becomes stronger and the other orientations become weaker after RTA for 100 seconds, as shown in Figure 2(C); (3) only the (220) orientation can be seen from XRD spectra after 180 seconds of RTA treatment under 1100℃, as shown in Figure 2(D). This result indicates that the amorphous silicon films has been highly crystallized with (220) preferred orientation after RTA treating at 1100℃, for 180 seconds.

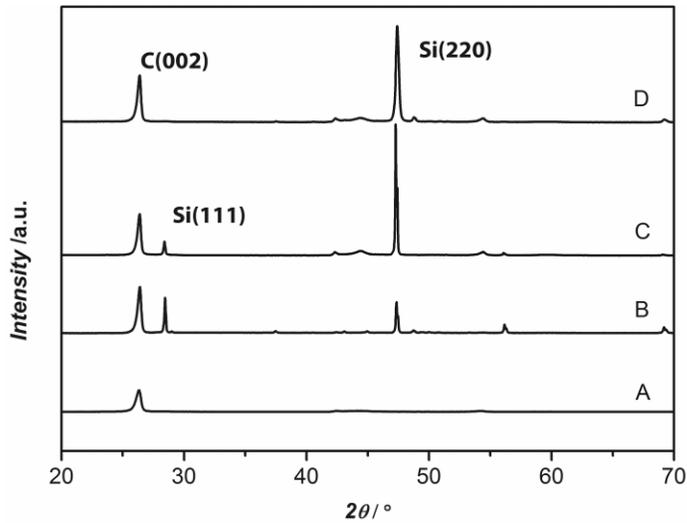

**Figure 2**. X-ray diffraction spectra of (A) amorphous silicon films, (B) CTA treated for 50 hours at 1100℃, (C) RTA treated for 100 seconds at 1100℃, and (D) RTA treated for 180 seconds at 1100℃.

Raman measurements are carried out as well for verifying the crystallization of amorphous silicon films. Raman spectra, shown in Figure 3, confirm further that the sample used in Figure 2(A) is amorphous (black line in Figure 3), and the sample used in Figure 2(D) is highly crystallized with preferred orientation after RTA treated for 180 seconds (red line in Figure 3,) because whose Raman spectrum is quite match that of a mono-crystalline silicon wafer (blue line in Figure 3). This result indicates that the amorphous silicon films has been highly crystallized to quasi-mono crystalline after RTA treating at 1100℃, for 180 seconds.

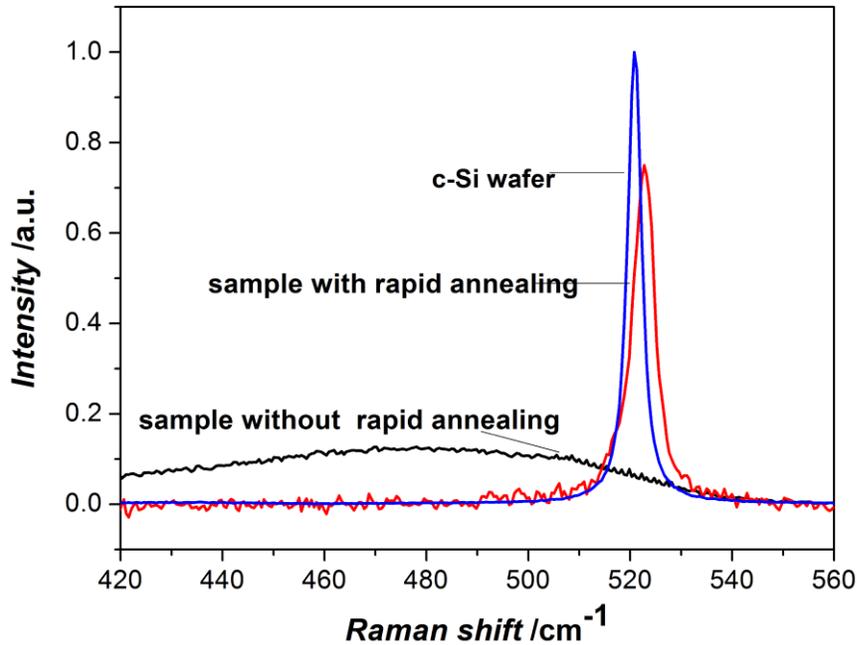

**Figure 3**. Raman spectra of an amorphous silicon film (black line), after RTA treated for 180 seconds (red line), and a comparing mono-crystalline silicon wafer (blue line).

**Analyses of orientation preferring**

It can be understood by theory of least-energy that the CTA crystallized silicon films behave (111) superior orientation as shown in Figure 2(B). Because it has been verified both theoretically and experimentally that the specific free surface energy of (111) plane is the smallest among all the other crystalline planes, as show in Table 1.[25,26] For the RTA crystallization, note that the thermal energy at 1100 ℃ is $E_T=kT=0.121eV$ (k is Boltzmann constant), the wave lengths of tungsten-halogen lamp arrange from 500 nm to 3000 nm, and the peak wave length is around 1000 nm, or the energy of photons near the peak radiation is $E_r=h\nu=1.241eV$. Obviously, $E_r$ is much greater than $E_T$. Therefore, photons may play more important role in the crystallization of amorphous silicon films than heat. How do photons affect on atoms composing amorphous silicon films and/or locating at the lattice points of crystalline silicon films? A simple model is set up as follows for explaining the changes of orientations during the crystallization of amorphous silicon films.

Table 1. Surface free energy and lattice point density on projection of (*hkl*) planes of silicon

| Atomic planes (*hkl*) | (00l) | (011) | (111) |
|---|---|---|---|
| Calculated surface free energy[a] ($10^{-4}$ J cm$^{-2}$) | 2.142 | 1.515 | 1.237 |
| Experimental surface free energy[b] ($10^{-4}$ J cm$^{-2}$) | 2.130 | 1.510 | 1.230 |
| Lattice point density on projection ($a^{-2}$) | 8 | 5.657 | 6.158 |

[a]Reference [25]

[b]Reference [26]

Suppose that the light-induced crystallization begins from the surface of amorphous silicon films under the bombardments of photons. It could be imagined that there might be a threshold value of energy $E_G$ like forbidden energy gap $E_g$ for electrons. Only the photons with energies greater than $E_G$ may affect on statuses of silicon atoms. The threshold value of energy $E_G$ might be supposed as the bonding energy of crystalline silicon at the primary stage. Under the bombardments of photons, the silicon atoms rearrange obeying the rule of least-energy as (111) orientation crystalline silicon at the beginning. With the extending of bombarding time the orientations with denser density of atoms on their projection planes will accept more serious bombardments of photons. Only the orientation with the lowest density of atoms on its projection plane will be kept after enough bombarding duration. All the atoms locating along the other orientations will rearrange to the orientation with the lowest density of atoms on its projection

plane.The lattice points on the projection of some typical planes (100), (110), and (111) of crystalline silicon are shown in Figure 4(A), 4(B), and 4(C) respectively.

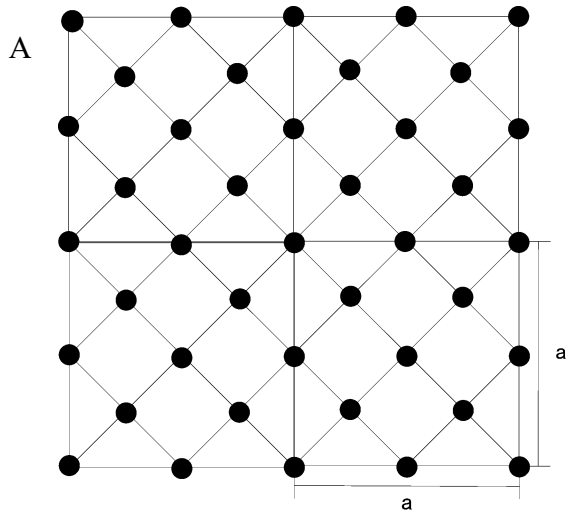

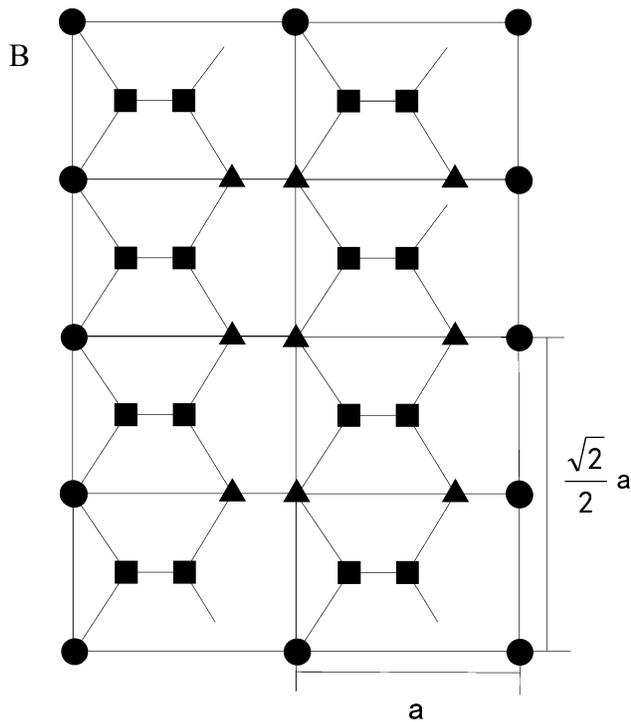

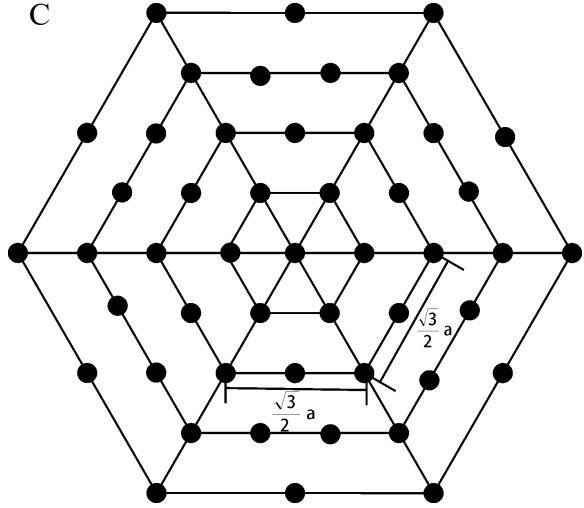

**Figure 4.** The lattice points on the projection of some typical planes (A) (001), (B) (011), and (C) (111) of crystalline silicon.

According to Figure 4, the densities of lattice points composed of $n^3$ unit cells on projection of planes (001), (011), and (111) of crystalline silicon are:

$$\rho(001) = N/S = (4 \times 1/4 + 4 \times 1/2 + 5)/a^2 = 8a^{-2} \tag{3}$$

$$\rho(011) = N/S = (8n^2 - n)/\sqrt{2}(na)^2 = ((8 - 1/n)/\sqrt{2})a^{-2} \tag{4}$$

$$\rho(111) = N/S = (12n^2 - 3n)/(9\sqrt{3}(na)^2/8) = ((32 - 8/n)\sqrt{3}/9)a^{-2} \tag{5}$$

Where $N$ is the number of atoms belongs to the projection area, $S$ is the area of the projection, and $a$ is the lattice parameter of silicon. Considering that the crystalline silicon is composed of huge unit cells, then n>>1, according to Equation (3), (4), and (5), the (110) orientation has the lowest density of atoms on its projection plane as listed in Table 1. Therefore it makes clear that why only (220) peak exists in the XRD diffraction spectra from enough RTA treated amorphous silicon film samples.

How does (111) orientation change to (110) orientation? The crystalline grains grow up gradually during the crystallization. The crystalline gains with (111) orientation will appear before the other orientations because of its smallest surface free energy. On the other hand, the effect of photons is much greater than heat on the crystallization of amorphous silicon films. As discussed above, the (110) orientation is the most stabilized. The angle between planes of (110) and (111) is 35.26°, as shown in Figure 5. Under the bombardments of photons, the crystalline grains with (111) orientation will turn into (110) orientation, because the angle between them is small and the latter is stable. In the other words, the orientation of crystalline grains can only be changed during their growing up by bombardments of photons. That is why a (111) orientation mono-crystalline silicon wafer cannot be changed to (110) orientation by RTA treatments.

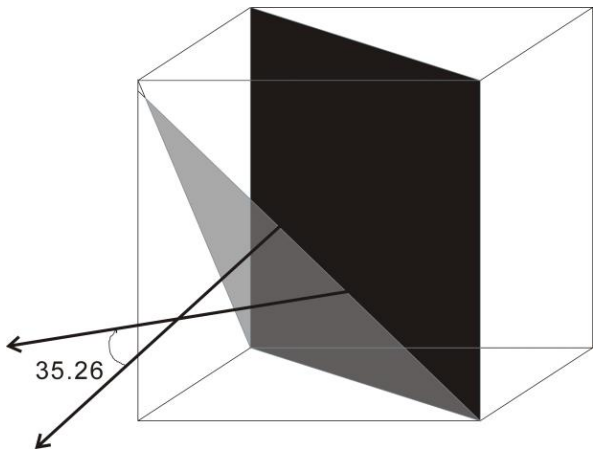

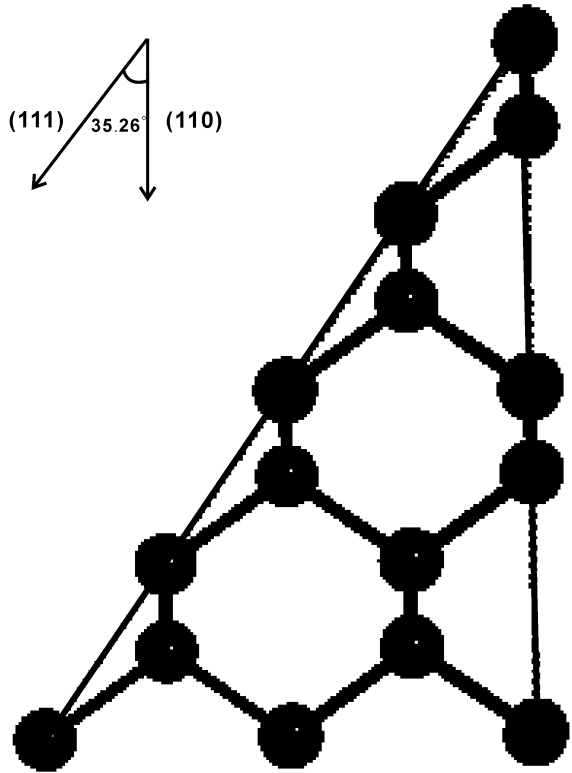

**Figure 5.** Angle between planes of (110) and (111) (A), and crystalline grain of silicon (B).

**Conclusion**

In summary, the crystallization of amorphous silicon films by RTA behaves (220) preferred orientation. If RTA duration is long enough, except (220) orientation, all the other orientations become weaker or disappear in x-ray diffraction spectra. The effect of photons on atoms take charge of the (220) preferred orientation. These phenomena indicate that statuses of silicon atoms may be changed by photons radiated from tungsten-halogen lamps.

These results can also be used to interpret the phenomena of laser ablation and irradiation-induced melting of metals. The (110) orientation preferred crystalline silicon films have important potential usage in fabricating radial solar cells,[27-30] because it is easier to make deep radial p-n junction on (110) orientation crystalline silicon films.

## ASSOCIATED CONTENT

**Supporting Information.** Calculation method for densities of atoms on projection planes

## ACKNOWLEDGMENT

Supported by Beijing Natural Science Foundation（2151004）

**Table of Contents**

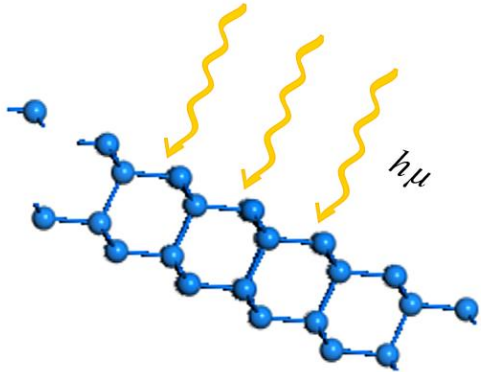

The crystallization of amorphous silicon films by RTA behaves (220) preferred orientation, and the effect of photons on atoms take charge of it. These phenomena indicate that statuses of silicon atoms may be changed by photons radiated from tungsten-halogen lamps